\documentclass[prb,twocolumn,amssymb,showpacs]{revtex4}
\usepackage{graphicx,amsmath,mathrsfs}
\begin{document}

\title{Half quantum spin Hall effect on the surface of weak topological insulators}

\author{Chao-Xing Liu$^{1,2}$, Xiao-Liang Qi$^3$, Shou-Cheng Zhang$^3$ }

\affiliation{ $^1$Institute for Theoretical Physics and Astrophysics,
University of W$\ddot{u}$rzburg, 97074 W$\ddot{u}$rzburg, Germany;\\
 $^2$Physikalisches Institut (EP3), University of W$\ddot{u}$rzburg, 97074 W$\ddot{u}$rzburg, Germany; \\
   $^{3}$Department of Physics, McCullough Building, Stanford University,
    Stanford, CA 94305-4045   }

\date\today

\begin{abstract}
We investigate interaction effects in three dimensional weak topological insulators (TI) with an even number of Dirac cones on the surface. We find that the surface states can be gapped by a surface charge density wave (CDW) order without breaking the time-reversal symmetry. In this sense, time-reversal symmetry alone can not robustly protect the weak TI state in the presence of interactions. If the translational symmetry is additionally imposed in the bulk, a topologically non-trivial weak TI state can be obtained with helical edge states on the CDW domain walls. In other words, a CDW domain wall on the surface is topologically equivalent to the edge of a two-dimensional quantum spin Hall insulator. Therefore, the surface state of a weak topological insulator with translation symmetry breaking on the surface has a ``half quantum spin Hall effect", in the same way that the surface state of a strong topological insulator with time-reversal symmetry breaking on the surface has a ``half quantum Hall effect". The on-site and nearest neighbor interactions are investigated in the mean field level and the phase diagram for the surface states of weak topological insulators is obtained.
\end{abstract}

\pacs{73.20.-r, 72.25.Mk, 71.45.Lr }

\maketitle

\section{Introduction}
Topological insulators (TIs) have recently attacted a lot of interest in the condensed matter physics community
due to its fundamental novelty and potential application\cite{qi2010,moore2010,hasan2010,qi2010a}.
Unlike normal insulators, TIs possess gapless modes with linear Dirac dispersion
at the edge or surface as a direct physical consequence of the bulk topology and the time reversal symmetry.
TIs have been realized in both two dimension, such as HgTe quantum wells\cite{bernevig2006d,koenig2007},
and three dimension, including Bi$_{1-x}$Sb$_x$\cite{hsieh2009}, Bi$_2$Se$_3$ family\cite{zhang2009,xia2009,chen2009} and 
TlBiSe$_2$ family\cite{yan2010,chen2010,kuroda2010,sato2010,lin2010}.
Within the non-interacting band theory, it is proposed that the TIs can be further classified into two classes\cite{fu2007a,moore2007,fu2007b},
one is the strong TI, which possesses odd number of Dirac cones at the surface
and has been realized in experiment as mentioned above, while the other one is the so-called weak TI,
which has even number of Dirac cones. The weak TI is topologically equivalent to a stack of two dimensional
TI (quantum spin Hall insulator) and is unstable against the translation breaking term\cite{fu2007a}.
It is also proposed that one dimensional helical liquid can exist along the dislocation line of the lattice in a weak TI,
which is believed to be protected by time reversal and stable against the weak disorder\cite{ran2009,imura2011}.
Recently, it is predicted that weak TI surface states are stable under time-reversal-invariant disorder\cite{ringel2011,mong2011}.

In the presence of interactions, the strong TI can be defined in terms of the topological field theory\cite{qi2008}
with a quantized axion angle of $\theta=\pi$. Since the bulk axion topological term can be expressed in terms of the
Chern-Simons term on the surface, the axion angle of $\theta=\pi$ implies a ``half quantum Hall effect" on the surface,
when time reversal symmetry is broken on the surface but preserved in the bulk. The half quantum Hall effect is
responsible for the topological magneto-electric effect of the strong TI\cite{qi2008}. The time reversal symmetry breaking
on the surface can be realized by Ising type magnetic moments oriented along the surface normal. A single chiral mode exists on the domain 
wall of the magnetic moments. Strong TI can be generally defined this way as long as the time reversal symmetry is
present inside the bulk, without specific reference to the translational symmetry of the lattice and the periodic band structure.
However, weak TI is not generally robust when only time reversal symmetry is present in the bulk, in fact we shall 
show that it is possible to open up a gap on the surface without breaking the time reversal symmetry. Therefore, lattice
translation symmetry must also be imposed in the bulk for the weak TI to be topologically well defined. In this case, it is
natural to ask what is the physical effect associated with breaking the translational symmetry on the surface of a weak TI.
In this paper, we investigate interaction effects in weak TI by first writing a minimal four band model for weak TIs.
Then we show that the weak TI phase, possessing two Dirac cones at the $(0,\pi)$ and $(\pi,0)$
of the surface Brillioun zone respectively, can be obtained by properly tuning the parameters of the four band model.
The surface staggered potential ($(\pi,\pi)$ charge density wave) can induce the scattering between the two Dirac cones
and gap the surface states without breaking time reversal symmetry. Interestingly, a helical liquid, identical to 
the edge states of the 2D topological quantum spin Hall insulator\cite{bernevig2006d,koenig2007},  emerges at the domain
wall for the staggered potential. Consequently, the gaped surface state of weak TI can be regarded as a half quantum spin Hall state,
in analogy to the half quantum Hall state for the gaped surface state of strong topological insulators.
We propose such a ``half quantum spin Hall effect" as a robust defining feature of weak TI. Since the quantum spin Hall state is stable upon electron interaction, so is the half quantum spin Hall effect. As long as the translation symmetry in the bulk is preserved, the weak TI is a well-defined topological state defined by the helical liquid on the domain wall of a surface CDW order. To find the condition for the surface CDW order, we further investigate the effect of repulsive interaction for the surface states of weak TIs. Our results show that the surface CDW state with half quantum spin hall effect is realized when nearest neighbor repulsion dominates the on-site repulsion, while a ferromagnetic quantum anomalous Hall state is induced in the opposite limit.
We find the surprising result that the antiferromagnetic order is not favored by the on-site repulsion on the surface of a weak TI, in sharp contrast to
conventional 2D systems.

\section{Half quantum spin Hall state}
The four band model introduced by Zhang {\it et al} has been successfully used to describe the strong topological insulator
of Bi$_2$Se$_3$ family of materials\cite{zhang2009}. For simplicity, we use the tight-binding regularization
to re-write the four band model in the cubic lattice
\begin{eqnarray}
	&&\hat{H}=\sum_{\bold{k}}\hat{\psi}^{\dag}(\bold{k})H({\bf k})\hat\psi({\bf k})\label{eq:HQSH_H0}\nonumber\\
	&&H(\bold{k})=\mathcal{M}(\bold{k})\Gamma_5+A\sum_{i=1,2,3}\Gamma_i\sin k_i
\end{eqnarray}
where
$\mathcal{M}(\bold{k})=M_0+6B-2B\sum_i\cos k_i$, $\Gamma_1=\sigma_1\otimes\tau_1$,
$\Gamma_2=\sigma_2\otimes\tau_1$, $\Gamma_3=\sigma_3\otimes\tau_1$, $\Gamma_4=1\otimes\tau_2$
and $\Gamma_5=1\otimes\tau_3$. Here $M_0$, $B$ and $A$ are model dependent parameters,
and for simplicity, we take $B>0$. We consider two orbitals, as well as spin, so that
the field operator
$\hat\psi(\bold{k})=[\hat{c}_{a,\uparrow},\hat{c}_{b,\uparrow},\hat{c}_{a,\downarrow},\hat{c}_{b,\downarrow}]^T$
(a and b are orbital indices). Pauli matrices $\sigma$ denote spin and $\tau$ denote orbital.
The model is assumed to be isotropic and preserve inversion symmetry $P=1\otimes\tau_3$.
Due to the inversion symmetry, the $Z_2$ topological nature of the system can be easily
extracted from the parities of the occupied states at the time reversal invariant points
according to Fu-Kane criterion\cite{fu2007a}. There are totally eight time reversal invariant
points in the whole 3D Brillioun zone (BZ), denoted as $\Lambda_i=(\kappa_1,\kappa_2,\kappa_3)$
($i=1\dots 8$) where $\kappa_{1,2,3}=0$ or $\pi$. The parities of the occupied bands can
be easily obtained from the sign of the mass term $\mathcal{M}({\bf k})$ for the above Hamiltonian (\ref{eq:HQSH_H0}),
denoted as $\delta_{\Lambda_i}=-sgn\left( \mathcal{M}({\bf k}=\Lambda_i) \right)$.
Then the strong TI $Z_2$ index is defined as\cite{fu2007a}
\begin{eqnarray}
	(-1)^{\nu_0}=\Pi_i\delta_{\Lambda_i}
	\label{eq:HQSH_nu0}
\end{eqnarray}
and three weak TI $Z_2$ indices are given by
\begin{eqnarray}
	(-1)^{\nu_k}=\Pi_{\kappa_k=\pi}\delta_{(\kappa_1,\kappa_2,\kappa_3)}.
	\label{eq:HQSH_nuk}
\end{eqnarray}
The strong TI index is determined by the product of the parities of all the eight $\Lambda_i$ point in BZ
while the weak TI index $\nu_k$ is given by the product of the parities of the four $\Lambda_i$ in
the $\kappa_k=\pi$ plane ($k=1,2,3$). With the above definition, the topological property of the Hamiltonian (\ref{eq:HQSH_H0}) is determined
by $(\nu_0;\nu_1\nu_2\nu_3)$, and we find the following different parameter regimes for the system.
When $M_0>0$ or $M_0<-12B$, the system is a trivial insulator with the $Z_2$ index
$(\nu_0;\nu_1\nu_2\nu_3)=(0;000)$; when $0>M_0>-4B$ or $-8B>M_0>-12B$, the system
shows the strong TI phase with the $Z_2$ index $(1;000)$ or $(1;111)$;
when $-4B>M_0>-8B$, the system stays in the weak TI phase with $Z_2$ index
$(0;111)$. Since here we are only interested in the weak TI phase, we focus on the parameter
regime $-4B>M_0>-8B$.

\begin{figure}
    \begin{center}
        \includegraphics[width=3.5in]{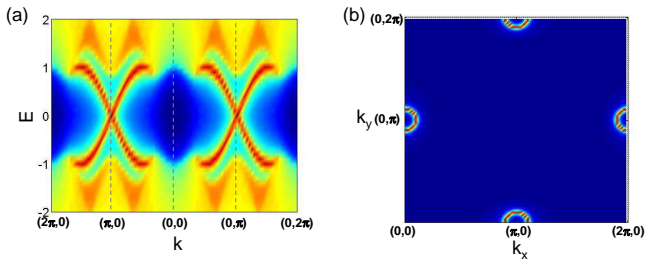}
    \end{center}
    \caption{ (a) The surface dispersion with two Dirac cones at $(0,\pi)$
    and $(\pi,0)$; (b) the local density of states at the surface BZ at the
    energy $E=0.3$. Here we take $A=1$, $B=1$ and $M_0=-5$. }
    \label{fig:surfaceLDOS}
\end{figure}

To investigate the surface state of weak TI, we consider the Hamiltonian (\ref{eq:HQSH_H0})
in a semi-infinite configuration (only $z<0$ region) with open boundary condition at $z=0$, so that the surface is normal to (100) direction (z direction).
It should be noted that the detailed form of surface state for weak TI will depend on the direction of the surface.
The surface states can be calculated numerically with the iterative Green function method\cite{dai2008}
and the obtained surface local density of state is shown in Fig \ref{fig:surfaceLDOS}.
In the regime $-4B>M_0>-8B$, it is found that there are two Dirac cones, located at
$\vec{Q}_1=(\pi,0)$ and $\vec{Q}_2=(0,\pi)$ of the surface BZ.
At the momentum $\vec{Q}_j$ (j=1,2), the eigen states of the Hamiltonian (\ref{eq:HQSH_H0})
can also be solved analytically with the eigen wave function given by
\begin{eqnarray}
	&&\Psi_{Q_j}^\sigma=\frac{1}{\sqrt{N}}e^{i\vec{Q}_j\cdot\vec{\rho}_n}\phi(n_z)\varphi_{0}^\sigma
	\label{eq:HQSH_wf1}
\end{eqnarray}
where $\vec{\rho}_n=(n_x,n_y)$, $\sigma$ denotes spin index and N is the normalization
factor. The spin part of the wave function is given by
\begin{eqnarray}
	&&\varphi_{0}^{\uparrow}=\left(
	\begin{array}{c}
		\xi_y\\0
	\end{array}
	\right)\qquad \varphi_{0}^{\downarrow}=\left(
	\begin{array}{c}
		0\\\xi^*_y
	\end{array}
	\right)
	\label{eq:HQSH_basiswf}
\end{eqnarray}
with $\xi_y=\frac{1}{\sqrt{2}}[i,1]^T$. The z-direction wave function takes
the form $\phi(n_z)\sim a_1\lambda_1^{n_z}+a_2\lambda_2^{n_z}$ where $\lambda_{1,2}$ are
the two solutions of the equation
\begin{eqnarray}
	\frac{A}{2}(\lambda-\lambda^{-1})+M_0+6B-B(\lambda+\lambda^{-1})=0.
	\label{eq:HQSH_lambda}
\end{eqnarray}
It is found that only in the regime $-4B>M_0>-8B$, $\lambda_{1,2}$ will satisfy
$|\lambda_{1,2}|>1$, hence a normalizable wave function ($z\rightarrow-\infty$ limit) exists,
which agrees with the parity criterion discussed above. With the eigen wave function of the surface states,
we project the Hamiltonian (\ref{eq:HQSH_H0}) into the sub-space spanned by $\Psi_{Q_j}^\sigma$, which
can be done by expanding the operator $\hat{\psi}(\vec{n})$ as
\begin{eqnarray}
	\hat{\psi}(\vec{n})\approx \sum_{j,\sigma} \Psi_{Q_j}^\sigma(\vec{\rho}_n,n_z)
	\hat{c}_{j,\sigma}(\vec{\rho}_n)
	\label{eq:HQSH_expan}
\end{eqnarray}
where the new operator $\hat{c}_{j,\sigma}(\vec{\rho}_n)$ is assumed to slowly
vary with $\vec{\rho}_n$. Substituting (\ref{eq:HQSH_expan}) into (\ref{eq:HQSH_H0}),
we find the effective Hamiltonian for the surface state
\begin{eqnarray}
	\hat{H}_{sur}
        =A\int d\vec{\rho}\Psi^\dag(\vec{\rho}) \left( k_x\sigma_2\otimes\tau_3
	+k_y\sigma_1\otimes\tau_3 \right)\Psi(\vec{\rho}),
	\label{eq:HQSH_Hsur}
\end{eqnarray}
with $\Psi^\dag(\vec{r})=(c^\dag_{1,\uparrow},c^\dag_{1,\downarrow},
c^{\dag}_{2,\uparrow},c^\dag_{2,\downarrow})$. Here $\sigma$ and $\tau$ denote Pauli matrices
in spin and the two Dirac cones at $\vec{Q}_{1,2}$ of the surface BZ.
This Hamiltonian satisfies time reversal symmetry $T=\Theta K$ with $\Theta=i\sigma_y\otimes 1$
and $K$ complex conjugation.

\begin{figure}
    \begin{center}
        \includegraphics[width=3.5in]{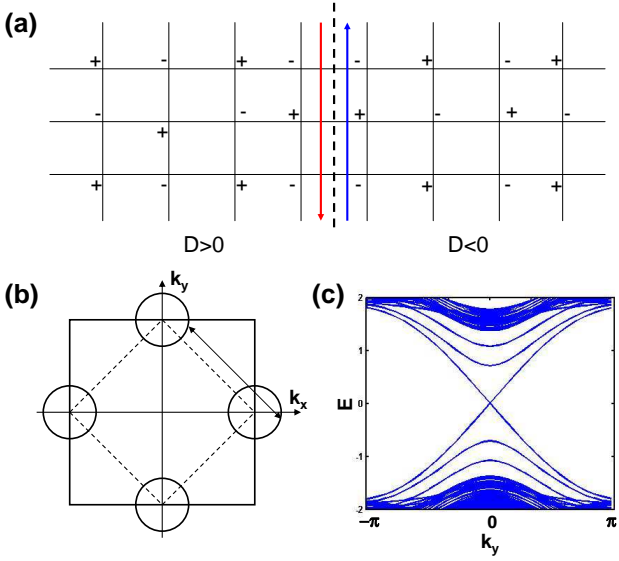}
    \end{center}
    \caption{ (a) Two possible staggerd potentials at the surface of weak TI.
    The dashed line shows the position of the domain wall of the staggered potential with a helical
    liquid along the domain wall. (b) The surface BZ and the reduced surface BZ (dashed line) when
    there is staggered potential. (d) The band dispersion of the system when there is
    a staggered potential domain wall. }
    \label{fig:HalfQSH}
\end{figure}

The next question is which kind of scattering can open a gap for the surface states
without breaking time reversal. With a single Dirac cone, it is found that
the two components are related to each other by time reversal symmetry, therefore
it is not possible to open a gap without breaking time reversal by the scattering
within a single Dirac cone. Consequently, we need to consider the scattering between
two Dirac cones, as shown in Fig \ref{fig:HalfQSH} (b). The only mass term which can
gap both Dirac cones while preserving time reversal symmetry is
\begin{eqnarray}
	&&H_1=D\int d^2r\Psi^\dag 1\otimes\tau_1\Psi.
	\label{eq:HQSH_mass}
\end{eqnarray}
To identify the physical meaning of $H_1$, we can re-write this term in the lattice model
and obtain the form $H_1=D\sum_n(-1)^{n_x-n_y}\hat{c}^\dag(\vec{n})\hat{c}(\vec{n})$, which is
nothing but the staggered potential (or $(\pi,\pi)$ charge density wave) at the surface.
The staggered potential breaks translation symmetry, enlarges the lattice unit-cell
and reduces the surface BZ. In the new surface BZ, as shown by the dashed line in Fig \ref{fig:HalfQSH} (b),
two Dirac cones can be related to each other by the reciprocal lattice vector,
therefore they actually correspond to the same point and can be gapped in the reduced BZ. Thus the weak TI requires
the protection of translation symmetry and can be adiabatically
connected to the normal insulator if the translation symmetry is broken.

\begin{figure}
    \begin{center}
        \includegraphics[width=3in]{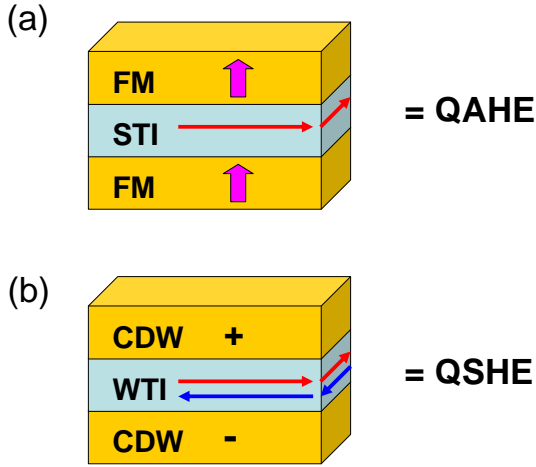}
    \end{center}
    \caption{ (a) Schematic plotting of the quantum anomalous Hall insulator with Hall conductance $e^2/h$ formed by a strong topological insulator sandwiched by
    two ferromagnetic materials at two surfaces, therefore one surface contributes half quantum Hall conductance.  (b) In contrast, when the weak topological insulator
    is sandwiched by two materials with opposite charge density wave at the two surfaces, a quantum spin Hall insulator is formed. Consequently, we can regard
    one surface as half quantum spin Hall state. }
    \label{fig:HalfQAHQSH}
\end{figure}

However there is still a non-trivial physical consequence for the gapped system.
For the surface staggered potential, there are two possiblilities, as shown
by the left side and right side of Fig \ref{fig:HalfQSH} (a),
which are related to each other by a shift of one lattice constant and
correspond to the mass term $D>0$ and $D<0$ in the Hamiltonian (\ref{eq:HQSH_mass}), respectively.
Now the surface states are described by Dirac equations, and when a mass domain wall forms
between two possible staggered potentials, zero modes are expected to emerge at the domain wall,
as shown in Fig \ref{fig:HalfQSH} (a). To see this, we consider a mass domain wall along x direction,
namely $D(x)=D_0>0$ when $x>0$ and $D(x)=-D_0<0$ when $x<0$. At $k_y=0$ point, the eigen equation for the zero modes
(with zero eigen-energy) reads $\partial_x\psi(x)=D(x)(\sigma_2\otimes\tau_2)\psi(x)$, which leads to
the normalizable domain wall solution $A\psi_s(x)=\frac{1}{\sqrt{N}}e^{-D_0|x|/A}\xi_{s-}$ ($s=\pm$), where $N$ is normalization factor
and $\xi_{st}$ is the eigen-state of $(\sigma_2\otimes\tau_2)$ with $(\sigma_2\otimes\tau_2)\xi_{st}=t\xi_{st}$.
$s=\pm$ denotes the additional degeneracy for the solution $\psi_s(x)$ and because $[\sigma_1\otimes\tau_3,\sigma_2\otimes\tau_2]=0$,
we can take $\xi_{st}$ also as the eigen-state of $\sigma_1\otimes\tau_3$ with $(\sigma_1\otimes\tau_3)\xi_{st}=s\xi_{st}$.
Because $T(\sigma_1\otimes\tau_3)T^{-1}=-(\sigma_1\otimes\tau_3)$, $(\sigma_1\otimes\tau_3)T\xi_{st}=-T(\sigma_1\otimes\tau_3)\xi_{st}
=-sT\xi_{st}$, which indicates that $\xi_{+-}$ and $\xi_{--}$ are two Kramers' partners. We can project the Hamiltonian (\ref{eq:HQSH_Hsur})
with nonzero $k_y$ into the subspace spanned by $\psi_\pm(x)$ and obtain the effective Hamiltonian
$H_{1D}=A\sum_{k_y} \left( \hat{c}_+^\dag k_y\hat{c}_+- \hat{c}_-^\dag k_y\hat{c}_-\right)$
where $\hat{c}_\pm$ and $\hat{c}_\pm^\dag$ denote the annihilation and creation operator for the zero modes $\psi_\pm(x)$.
From the form of the effective Hamiltonian, the zero mode solution at the mass domain wall is exactly the helical liquid, consisting of only one left mover and one right mover,
which are related to each other by time reversal, identical to the edge states discovered in the 2D quantum spin Hall state\cite{bernevig2006d,koenig2007}.
A direct numerical calculation of the surface states with the mass domain wall exactly gives the above physical picture,
as shown in Fig \ref{fig:HalfQSH} (c). Here we can furthermore make the analogy to the surface state of
the strong TI.  When the top and bottom surfaces of a strong TI slab are covered by magnetic materials
with the magnetization direction as shown in Fig. 14 of Ref. \cite{qi2008} and reproduced here in Fig \ref{fig:HalfQAHQSH} (a), the system has one chiral edge mode and exhibit the Hall conductance
of $\frac{e^2}{h}$. Since two surfaces contribute equally to the $\frac{e^2}{h}$ Hall conductance, each surface is expected to contribute
half of the Hall conductance $\frac{e^2}{2h}$, leading to the so-call half quantum Hall effect\cite{fu2007a,qi2008}.
Here similarly, consider the weak TI slab sandwiched by two $(\pi,\pi)$ charge density wave materials, then the system will become a quantum
spin Hall insulator with a helical mode along the edge, as shown in Fig \ref{fig:HalfQAHQSH} (b).
Again the top and bottom surfaces contribute equally to the quantum spin Hall insulator, consequently we can regard one surface state as a half quantum spin Hall state.


\section{Interaction effect}
In this section we hope to investigate the effect of the interaction
on the surface states of weak TI within the mean field approximation.
Here the on-site and nearest neighbour interaction are taken into account,
with the form
\begin{eqnarray}
	&&\hat{H}_{U}=U\sum_{\vec{n},\eta}\hat{c}^\dag_{\eta\uparrow}(\vec{n})\hat{c}_{\eta\uparrow}(\vec{n})
	\hat{c}^\dag_{\eta\downarrow}(\vec{n})\hat{c}_{\eta\downarrow}(\vec{n})\nonumber\\
        &&\hat{H}_{V}=V\sum_{\langle i,j\rangle}\hat{n}_i\hat{n}_j,\qquad
	\hat{n}_i=\sum_{\eta\sigma}\hat{c}^\dag_{\eta\sigma}(\vec{i})\hat{c}_{\eta\sigma}(\vec{i}).
\end{eqnarray}
where $\eta=a,b$ denotes orbitals and $\sigma=\uparrow,\downarrow$ denotes spin.
$\hat{H}_U$ and $\hat{H}_V$ can also be projected into the sub-space of the surface states
by the expansion (\ref{eq:HQSH_expan}) for the operator $\hat{\psi}(\vec{n})$, and the obtained
Hamiltonian is given by
\begin{eqnarray}
	&&\hat{H}_{U}=\tilde{U}\int d^2r\sum_{i=0,1}\left[ (\Psi^\dag\sigma_0\otimes\tau_i\Psi)
	(\Psi^\dag\sigma_0\otimes\tau_i\Psi)\right.\nonumber\\
	&&\left.-(\Psi^\dag\sigma_3\otimes\tau_i\Psi)(\Psi^\dag\sigma_3\otimes\tau_i\Psi)
	\right]\label{eq:IE_HUV1}\\
	&&\hat{H}_V=\int d^2r\left[ \tilde{V}_1 \hat{\Psi}^\dag(\vec{r})\sigma_0\otimes\tau_0\hat{\Psi}(\vec{r})\hat{\Psi}^\dag(\vec{r})
	\sigma_0\otimes\tau_0\hat{\Psi}(\vec{r})\right.\nonumber\\
	&&\left.-\tilde{V}_2\hat{\Psi}^\dag(\vec{r})\sigma_0\otimes\tau_1\hat{\Psi}(\vec{r})
	\hat{\Psi}^\dag(\vec{r})\sigma_0\otimes\tau_1\hat{\Psi}(\vec{r})\right]
	\label{eq:IE_HUV2}
\end{eqnarray}
where $\tilde{U}=\frac{UN_1}{8}$, $\tilde{V}_1=V(N_2+2N_1)$, $\tilde{V}_2=V(2N_1-N_2)$,
$N_1=\sum_{n_z}|\phi(n_z)|^4$ and $N_2=\sum_{i_z}|\phi(i_z)\phi(i_z+1)|^2$.

We can treat the interaction Hamiltonian (\ref{eq:IE_HUV1}) and (\ref{eq:IE_HUV2}) in the mean field level
and need to take into account the following order parameters
\begin{eqnarray}
	&&D=\langle \hat{\Psi}^\dag\sigma_0\otimes\tau_1\hat{\Psi}\rangle,\qquad
	\vec{m}=\langle \hat{\Psi}^\dag\vec{\sigma}\otimes\tau_0\hat{\Psi}\rangle,\nonumber\\
	&&\vec{S}=\langle \hat{\Psi}^\dag\vec{\sigma}\otimes\tau_1\hat{\Psi}\rangle.
	\label{eq:IE_oderparameter}
\end{eqnarray}
Here $D$, $\vec{m}$ and $\vec{S}$ are the order parameter of $(\pi,\pi)$ charge density wave, ferromagnetism
and anti-ferromagnetism, respectively. The mean field Hamiltonian is given by
\begin{eqnarray}
	&&\hat{H}_{U}=\tilde{U}\int d^2r\left[ 2n(\Psi^\dag\sigma_0\otimes\tau_0\Psi)-n^2\right.\nonumber\\
	&&+2D(\Psi^\dag\sigma_0\otimes\tau_1\Psi)
	-D^2-2m_3(\Psi^\dag\sigma_3\otimes\tau_0\Psi)\nonumber\\
	&&\left.+m^2_3-2S_3(\Psi^\dag\sigma_3\otimes\tau_1\Psi)+S^2_3\right]\\
	&&\hat{H}_{V}=\int d^2r\left\{ \tilde{V}_1\left[ 2n(\Psi^\dag\sigma_0\otimes\tau_0\Psi)-n^2 \right]\right.\nonumber\\
	&&\left.-\tilde{V}_2\left[ 2D(\Psi^\dag\sigma_0\otimes\tau_1\Psi)-D^2 \right]\right\},
	\label{eq:IE_meanfieldH}
\end{eqnarray}
with the density $n=\langle \hat{\Psi}^\dag\hat{\Psi}\rangle$.

The phase diagram of the system can be obtained by minimizing the total free energy
\begin{eqnarray}
	\hat{F}=\hat{H}_{sur}+\hat{H}_U+\hat{H}_V-\mu\int d^2r\hat{\Psi}^\dag\hat{\Psi}
	\label{eq:IE_freeenergy}
\end{eqnarray}
The last term gives the chemical potential which determines the density $n$,
and when the chemical potential is at the Dirac point of the surface states,
the density $n$ is taken to be zero.

Next let's analyze the possible non-trivial phase induced by the interaction Hamiltonian
$\hat{H}_U$ and $\hat{H}_V$. First we only consider the nearest neighbour interaction $\hat{H}_V$.
From Eq. (\ref{eq:IE_meanfieldH}), we can see that $\tilde{V}_2$ term prefers
$(\pi,\pi)$ charge density wave. By minimizing the free energy,
we obtain the self-consistent equantion
\begin{eqnarray}
	\frac{1}{\tilde{V}_2}=\frac{2}{\pi}\left( \sqrt{A^2\Lambda^2+(2\tilde{V}_2D)^2}-\sqrt{2\pi nA^2+(2\tilde{V}_2D)^2} \right).
\end{eqnarray}
with the high energy cut-off $\Lambda$. For the above equation, a self-consistent solution
only exists when $\tilde{V}_2>\tilde{V}_{2c}=\frac{A\pi}{2(\Lambda-\sqrt{2\pi n})}$.
When this condition is satisfied, there is $(\pi,\pi)$ charge density wave in the system,
which induces the half quantum spin Hall state.

For $\hat{H}_U$, the situation is a little more complicated since there are three different order parameters:
ferromagnetism $m_3$, anti-ferromagnetism $S_3$ and charge density wave $D$. Charge density wave $D$
has been discussed above and here we focus on the competition between the ferromagnetic order $m_3$ and
anti-ferromagnetic order $S_3$. Interestingly, we find that the ferromagnetic order $m_3$
will induce a gap of the system but the anti-ferromagnetic order $S_3$ can not.
Consequently we expect that the ferromagnetic order will win, which is different from the usual two dimensional electron gas
where the on-site Hubbard interaction will induce the anti-ferromagnetic order. Indeed, the minimum of the
total free energy is determined by the self-consistent equations
\begin{eqnarray}
	\tilde{U}=\frac{\pi}{2\Lambda^2}\left( \sqrt{A^2\Lambda^2+(2\tilde{U}m_3)^2}+2\tilde{U}m_3 \right),
	S_3=0.
\end{eqnarray}
Non-zero ferromagnetic order requires $\tilde{U}>\tilde{U}_c=\frac{\pi A}{2\Lambda}$.
In such case, the time reversal symmetry is spontenously broken and the ferromagnetic order
open a gap for each Dirac cone separately. Since each Dirac cone will contribute half quantized
Hall conductance when it is gapped, the surface state has Hall conductance $\frac{e^2}{h}$.

\begin{figure}
    \begin{center}
        \includegraphics[width=3in]{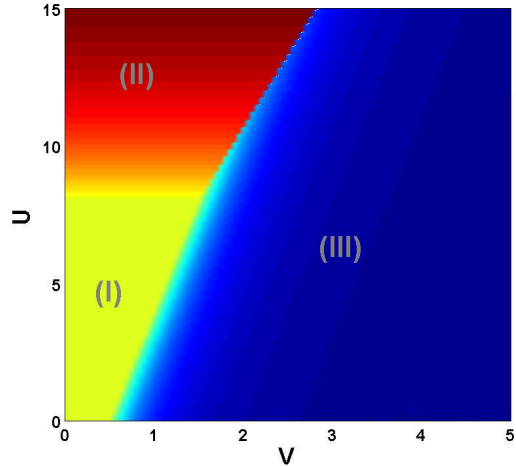}
    \end{center}
    \caption{ The phase diagram for the interacting surface states of weak TI. Here we take $N_1=1$ and $N_2=0$.
    $U$ and $V$ are in the unit of $\frac{\pi A}{2\Lambda}$.
    (I) semi-metal regime; (II) ferromagnetic and quantum anomalous Hall regime; (III) charge density wave and
    half quantum spin Hall regime.  }
    \label{fig:phase}
\end{figure}

Now let's discuss the coexistence of the on-site and nearest neighbour interaction,
where the ferromagnetic order $m_3$ will compete with the charge density wave order $D$.
In such case, we only consider the competition of these two orders and obtain
\begin{eqnarray}
	&&\frac{2\pi m_3}{\Lambda^2}=\sum_{t=\pm}\frac{2m_3\tilde{U}+2tD(\tilde{V}_2-\tilde{U})}
	{f(m_3,D)}, \\
	&&\frac{2\pi D}{\Lambda^2}=\sum_t\frac{2tm_3\tilde{U}+2D(\tilde{V}_2-\tilde{U})}
	{f(m_3,D)}.
\end{eqnarray}
with $f(m_3,D)=\sqrt{A^2\Lambda^2+(2m_3\tilde{U}+2tD(\tilde{V}_2-\tilde{U})^2)}+|2m_3\tilde{U}+2tD(\tilde{V}_2-\tilde{U})|$.
The equations can be solved numerically and the phase diagram is shown in Fig. \ref{fig:phase}.
The system remains in the semi-metal phase for the small U and V. When U is increased, there is a continous
phase transition to the ferromagnetic order with quantum anomalous Hall effct, while when V is increased,
there is a phase transition to the charge density wave order with half quantum spin Hall effect. When U and V
coexist, there is a first order phase transition. This picture is quite similar to the interaction effect
in the honeycomb lattice\cite{raghu2008}, and the non-trivial thing here is that the charge density wave can
induce half quantum spin Hall effect.

\section{Discussion and conclusions}
As a summary, we investigate interaction effects for the surface states
of the weak TI and the related phase diagram. We show that the weak TI can be topologically
robust if we supplement time reversal symmetry with translational symmetry in the bulk.
Helical modes exist on the domain wall of CDW order on the surface and the associated
``half quantum spin Hall effect" of the surface can be used as a general defining 
physical property of the weak TI. This provides one example of that the symmetry breaking 
inducing non-trivial physical phenomena.
Recently, several papers\cite{ringel2011,mong2011} discuss the disorder effect of the weak TI and show that the weak TI is robust
against the time reversal invariant disorder. This is consistent with our results here because in their
case, the translation symmetry is not broken macroscopically after taking the average of the disorder,
while in our case, the charge density wave breaks the translation symmetry macroscopically which induces
a band gap for the surface state.
Moreover, we conjecture that the weak anti-localization for the surface state of weak TI under the time reversal invariant disorder can also be understood as a consequence of the half quantum spin Hall effect we studied. Consider the clean surface with CDW order, and turn on non-magnetic impurities. The disorder will pin different CDW domains around different spatial points, leading to random CDW domain walls on the surface. There is a pair of helical edge states propagating along each domain wall. With random impurities, the area of the two types of CDW domains are always equal on average, so that the edge channels are always conducting due to percolation.

Finally, we remark on the possible material realization of the weak TI. The only known material exhibiting
non-trivial weak TI index is Bi$_{1-x}$Sb$_x$ alloy\cite{fu2007a,hsieh2008}, which also has non-trivial strong TI index. Consequently, if a charge density wave order is introduced on the surface of Bi$_{1-x}$Sb$_x$, the surface will become a half quantum spin Hall state coexisting with an additional massless Dirac cone from the strong TI index. The weak TI phase is also proposed recently for the 1/6 and 2/3 filling fractions
in the octahedron-decorated cubic lattice\cite{hou2011}. Search for new weak TI materials will be another interesting question for the future research. We also notice a recent paper with a related but different proposal that charge density wave can also induce quantum spin Hall effect in some 2D systems\cite{guo2011}.

\acknowledgments
We would like to thank G. Li, B. Trauzettel for the helpful discussion.
SCZ is supported by the NSF under grant numbers DMR-0904264 and the
Keck Foundation. XLQ is supported by the Alfred P. Sloan Foundation.
CXL acknowledge financial support by the Alexander von Humboldt Foundation of Germany.
CXL is grateful for the hospitality of the Institute for Advanced Study in Tsinghua University,
where the paper was completed.



%
\end{document}